\documentclass[10pt]{elsart} 
\usepackage{array}
\usepackage{oldlfont}
\usepackage{amssymb}
\usepackage{amsmath}
\journal{Physics Letters A}
\begin{document}
\begin{frontmatter}
\title{Upper limit on the number of bound states of the spinless Salpeter equation}
\author{Fabian Brau\thanksref{fnrs}}
\thanks[fnrs]{FNRS Postdoctoral Researcher}
\address{Service de Physique G\'en\'erale et de Physique des Particules El\'ementaires, Groupe de Physique Nucl\'eaire Th\'eorique, Universit\'e de Mons-Hainaut, Mons, Belgique}
\ead{fabian.brau@umh.ac.be}
\date{\today}

\begin{abstract}
We obtain, using the Birman-Schwinger method, upper limits on the total number of bound states and on the number of $\ell$-wave bound states of the semirelativistic spinless Salpeter equation. We also obtain a simple condition, in the ultrarelativistic case ($m=0$), for the existence of at least one $\ell$-wave bound states: $C(\ell,p/(p-1))$ $\int_0^{\infty}dr\, r^{p-1}\, |V^-(r)|^p\ge 1$, where $C(\ell,p/(p-1))$ is a known function of $\ell$ and $p>1$.
\end{abstract}

\end{frontmatter}

\section{Introduction}
\label{sec1}
The spinless Salpeter equation is a simple relativistic version of the Schr\"odinger equation which can be obtained, with some approximations, from the covariant Bethe-Salpeter equation \cite{salp51,grei94} and takes the form ($\hbar=c=1$)
\begin{equation}
\label{eq1}
\left[\sqrt{{\bf p}^2+m^2}+V({\bf r})\right]\, \Psi({\bf r})=M\, \Psi({\bf r}),
\end{equation}
where $m$ is the mass of the particle and $M$ is the mass of the eigenstate ($M=m+E$, $E$ is the binding energy).
We restrict our attention to time component vector potentials. This equation is generally used when kinetic relativistic effects cannot be neglected and when the particles under consideration are bosons or when the spin of the particles is neglected or is only taken into account via spin-dependent interactions. Despite its apparent complexity, this equation is often preferred to the Klein-Gordon equation. The equation (\ref{eq1}) appears, for example, in mesons and baryons spectroscopy in the context of potential models (see for example \cite{st80,go85,ca86,fu94,se97,br98,gl98,br02}, see also \cite{lu99}). 

Due to the pseudo-differential nature of the kinetic energy operator, few exact results are known about this equation. Most of these results have been obtained for a Coulomb potential (for example, upper and lower bounds on energy levels) \cite{he77,ca84,ha84,ma89,ra94}. Recently, upper and lower limits on energy levels have been obtained for some other particular interactions \cite{ha01a,ha01b}.

Conversely to the Schr\"odinger equation, for which a fairly large number of results giving both upper and lower limits on the number of bound states can be found in the literature (see for example \cite{ba52,bi61,sc61,ca65a,ca65b,ca65c,ch68,ma72,gl76,si76,ma77,li80,ch95a,ch95b,ch96,bl96,br03}), only one result, concerning the total number of bound states, is known for the spinless Salpeter equation \cite{daub83}. After recalling in Section \ref{sec2} a general method to obtain upper limits on the number of bound states due to Birman \cite{bi61} and Schwinger \cite{sc61}, we derive such limits for the spinless Salpeter equation in Section \ref{sec3}. The Section \ref{sec4} is devoted to test these limits. In Section \ref{sec5}, we show how a simple modification of the limits found in Section \ref{sec3} leads to upper limits on the number of bound states that lie below a given energy.

\section{The Birman-Schwinger method}
\label{sec2}

Birman \cite{bi61} and Schwinger \cite{sc61} have shown how to obtain an upper limit on the number of bound states once the Green function of the kinetic energy operator of a wave equation is known. In this section, we recall briefly the main line of the method; for more details see the original articles \cite{bi61,sc61}.

Let $T\left({\bf p}^2\right)$ be a general kinetic energy operator and let
\begin{equation}
\label{eq2}
\left[T\left({\bf p}^2\right)+V({\bf r})\right]\Psi({\bf r})=E\, \Psi({\bf r})
\end{equation}
be the wave equation, in three dimensions, that satisfy the wave function $\Psi({\bf r})$ (eigenstates) and where $E$ is the energy (eigenvalues). Let $G(\Delta)$ be the Green function of $T\left({\bf p}^2\right)$:
\begin{equation}
\label{eq3}
T\left({\bf p}^2\right)G(\Delta)=\delta^3({\bf \Delta}),
\end{equation}
where ${\bf \Delta}={\bf r}-{\bf r}'$, $\Delta=|{\bf \Delta}|$ and $\delta^3({\bf x})$ is the Dirac function. We can write (\ref{eq2}), with $E=0$, using this Green function as
\begin{equation}
\label{eq4}
\Psi({\bf r})=-\int d{\bf r}'\, G(\Delta)\, V({\bf r}')\, \Psi({\bf r}').
\end{equation}
Since the purpose of the method is to obtain an upper limit on the number of bound states, we can replace $V({\bf r})$ by $-|V^-({\bf r})|$ where $V^-({\bf r})$ is the negative part of the potential obtained by setting the positive part of the potential equal to zero. Indeed, a decrease of the potential in some region must lower the energies of the bound states and therefore cannot lessen their number. Moreover, we introduce the parameter $0<\lambda\le1$ by the substitution $|V^-({\bf r})|\rightarrow \lambda|V^-({\bf r})|$. As $\lambda$ increases from 0, we reach a critical value, $\lambda_1$, at which a bound state first appears with a vanishing binding energy, $E=0$. With further growth of $\lambda$, the energy of this state decreases until we reach a second critical  value, $\lambda_2$, at which a second bound state appears and so on. When $\lambda$ has attained the value unity and, $\lambda_N\le 1<\lambda_{N+1}$, there are $N$ bound states.

We now introduce, to obtain a symmetrical kernel, a new wave function as
\begin{equation}
\label{eq5}
\Phi({\bf r})=|V^-({\bf r})|^{1/2}\, \Psi({\bf r}).
\end{equation}
The equation (\ref{eq4}) becomes
\begin{equation}
\label{eq6}
\Phi({\bf r})=\lambda\int d{\bf r}'\, K({\bf r},{\bf r}')\, \Phi({\bf r}'),
\end{equation}
where $K({\bf r},{\bf r}')$ is given by
\begin{equation}
\label{eq7}
K({\bf r},{\bf r}')=|V^-({\bf r})|^{1/2}\, G(\Delta)\,|V^-({\bf r}')|^{1/2}.
\end{equation}
If the kernel is positive, we have $0<\lambda_1<\lambda_2<\cdots<\lambda_N\le 1$ and
$0<\lambda_k<\infty$ ($\lambda_k$ denotes each eigenvalue of (\ref{eq6})). It is well known that the trace of the iterated kernels equals the sum of the eigenvalues of the integral equation (\ref{eq6}) as follow
\begin{equation}
\label{eq8}
\sum_{k=1}^{\infty} \frac{1}{(\lambda_k)^n} = \int d{\bf r}\, K^{(n)}({\bf r},{\bf r}),
\end{equation}
where the iterated kernel $K^{(n)}({\bf s},{\bf t})$ is given by
\begin{subequations}
\label{eq9}
\begin{equation}
\label{eq9a}
K^{(n)}({\bf s},{\bf t}) = \int d{\bf u}\ K({\bf s},{\bf u})\, K^{(n-1)}({\bf u},{\bf t}),
\end{equation}
with 
\begin{equation}
\label{eq9b}
K^{(1)}({\bf s},{\bf t}) \equiv K({\bf s},{\bf t}),
\end{equation}
\end{subequations}
and $n=1,2,\ldots$. Now it is plain that the following inequalities hold
\begin{equation}
\label{eq10}
\sum_{k=1}^{\infty} \frac{1}{(\lambda_k)^n}\ge \sum_{k=1}^{N} \frac{1}{(\lambda_k)^n}>N,
\end{equation}
where $N$ is the number of bound states. From (\ref{eq8}), (\ref{eq9}) and (\ref{eq10}) we find that an upper limit on the total number of bound states of the wave equation (\ref{eq2}) is given by
\begin{equation}
\label{eq11}
N<\int d{\bf r}\, K^{(n)}({\bf r},{\bf r}).
\end{equation}

\section{Application to the spinless Salpeter equation}
\label{sec3}
\subsection{Upper limit on the total number of bound states}
\label{subsec3.1}

To obtain an upper limit on the total number of bound states of the spinless Salpeter equation, we need to calculate the Green function of the kinetic energy operator. Similar calculations have already been performed previously \cite{nick84,brau98}. In contrast to calculations found in \cite{brau98}, we need here to calculate the Green function of the following operator
\begin{equation}
\label{eq13}
T\left({\bf p}^2\right)=\sqrt{{\bf p}^2+m^2}-m.
\end{equation}
This is done by performing the integral
\begin{equation}
\label{eq14}
G(m,\Delta)=\frac{1}{(2\pi)^3}\int d{\bf p}\, \frac{\exp(-i\, {\bf p}.{\bf \Delta})}
{\sqrt{p^2+m^2}-m}.
\end{equation}
We find that
\begin{subequations}
\label{eq15}
\begin{equation}
\label{eq15a}
G(m,\Delta)=\frac{m}{4\pi \Delta}\left[1+\frac{2}{\pi}F(m\Delta)\right]\equiv 
\frac{m}{4\pi \Delta}\, H(m\Delta),
\end{equation}
with
\begin{equation}
\label{eq15b}
F(y)=\int_y^{\infty} \frac{dz}{z}\, K_1(z)+\frac{\pi}{2},
\end{equation}
\end{subequations}
where $K_{\nu}(y)$ is a modified Bessel function (see for example \cite[p. 374]{abra70}).
The zero-energy spinless Salpeter equation takes thus the form of the following integral equation
\begin{equation}
\label{eq16}
\Psi({\bf r})=\int d{\bf r}'\, G(m,\Delta)\, |V^-({\bf r}')|\, \Psi({\bf r}'),
\end{equation}
with $G(m,\Delta)$ given by (\ref{eq15}), and where we have replaced the potential by its negative part. This integral equation (\ref{eq16}) can be written with a symmetrical kernel:
\begin{subequations}
\label{eq17} 
\begin{equation}
\label{eq17a}
\Phi({\bf r})=\int d{\bf r}'\, K(m,{\bf r},{\bf r}')\, \Phi({\bf r}'),
\end{equation}
where
\begin{equation}
\label{eq17b}
K(m,{\bf r},{\bf r}')=|V^-({\bf r})|^{1/2} G(m,\Delta) |V^-({\bf r}')|^{1/2},
\end{equation}
\end{subequations}
and $\Phi({\bf r})$ given by (\ref{eq5}).
We can now use the Birman-Schwinger method, and in particular the relation (\ref{eq11}), with the integral equation (\ref{eq17}) to obtain the following upper limit on the total number of bound states of the spinless Salpeter equation
\begin{equation}
\label{eq17-a}
N<\frac{1}{\alpha^n}\int d{\bf r}_1 \ldots d{\bf r}_n\, |V^-({\bf r}_1)|\ldots |V^-({\bf r}_n)|\, G(m,\Delta_{12})\ldots
G(m,\Delta_{n1}),
\end{equation}
where $\Delta_{ij}=|{\bf r}_i-{\bf r}_j|$ and $n \ge 4$ (the integral diverges for smaller values of $n$). We have introduced in (\ref{eq17-a}) the parameter $\alpha$ which takes the value 1 respectively 2 for one respectively two (identical) particle problems.

Now, we need to calculate the Green function (\ref{eq15}) in a closed form, that is to say, to compute $F(y)$, see (\ref{eq15b}). An integration by part leads to
\begin{equation}
\label{eq17-b}
F(y)=K_1(y)+\frac{\pi}{2}-\int_y^{\infty} dz\, K_0(z).
\end{equation}
To obtain an upper limit for the number of bound states, a majorization of the kernel (\ref{eq17b}) is enough. Since the Bessel function $K_0(z)$ is positive for $0\le z <\infty$, and that its integration between 0 and $\infty$ is equal to $\pi/2$ (see \cite[p. 486]{abra70}), the integral in (\ref{eq17-b}) is not only positive but also small compared to the other terms of this equation. Indeed, in the region ($y\approx 0$) where this integral takes its maximal value ($\pi/2$), this quantity is still small compared to the value taken by the singular Bessel function $K_1(y)$. Thus the majorization obtained by replacing the integral in (\ref{eq17-b}) by 0, namely 
\begin{equation}
\label{eq17-c}
F(y)\le K_1(y)+\pi/2,
\end{equation}
should not spoil too much the upper limit as one can verify with an examination of the results of the tests performed in Section \ref{sec4}. Another majorization, which leads to a simpler kernel, is obtained by replacing the singular Bessel function $K_1(y)$ by $1/y$. This additional approximation is however exact in the case of a vanishing mass ($m=0$) since 
$m K_1(my)=1/y$ in this limit. The kernel is then given by
\begin{subequations}
\begin{eqnarray}
\label{eq17-d}
G(m,\Delta_{ij})&\le& \frac{m}{2\pi \Delta_{ij}}+\frac{m K_1(m\Delta_{ij})}{2\pi^2 \Delta_{ij}}=G^{(1)}(m,\Delta_{ij})\\
&\le& \frac{m}{2\pi \Delta_{ij}}+\frac{1}{2\pi^2 \Delta_{ij}^2}=G^{(2)}(m,\Delta_{ij}).
\end{eqnarray}
\end{subequations}
The upper limit (\ref{eq17-a}) can then be used with either $G^{(1)}(m,\Delta_{ij})$ or $G^{(2)}(m,\Delta_{ij})$. Of course, using the function $G^{(1)}(x,y)$ yields more stringent results than those obtained by using the function $G^{(2)}(x,y)$.

We will discuss more specifically the case of a central potential in the Section \ref{subsec3.2} but we already write here the upper limit (\ref{eq17-a}), using the kernel $G^{(2)}(x,y)$, for this class of potentials. In this way, we introduce some quantities which will be useful later.
Integrating over angular variables, the limit (\ref{eq17-a}) reads
\begin{subequations}
\label{eq17e}
\begin{eqnarray}
\label{eq17-e}
N<\sum_{\nu=0}^{\infty}(2\nu+1)\int_0^{\infty} dr_1 \ldots dr_n\, |V^-(r_1)|\ldots |V^-(r_n)| \times\nonumber \\ \times A_{\nu}(m,r_1,r_2)\ldots
A_{\nu}(m,r_n,r_1),
\end{eqnarray}
with 
\begin{equation}
\label{eq17-f}
A_{\nu}(m,x,y)=\frac{1}{\alpha}\frac{4\pi}{2\nu+1}\, xy\, a_{\nu}(m,x,y),
\end{equation}
and
\begin{eqnarray}
\label{eq17-g}
a_{\nu}(m,x,y)=&&\frac{m}{2\pi}\, r_{<}^{\nu}\,r_{>}^{-(\nu+1)}+\frac{1}{2\pi^{3/2}}\frac{\Gamma(\nu+1)}{\Gamma(\nu+1/2)}
\frac{(xy)^{\nu}}{(x+y)^{2(\nu+1)}}\nonumber \\ &&F\left(\nu+1,\nu+1,2(\nu+1),\frac{4xy}{(x+y)^2}\right),
\end{eqnarray}
\end{subequations}
where $r_<=\min[x,y]$, $r_>=\max[x,y]$ and $F(a,b,c,z)$ is the hypergeometric function.

A simpler upper limit can be obtained with additional approximations in the ultrarelativistic case ($m=0$). Using $n$ times ($n\ge 4$) the H\"older inequality we obtain
\begin{subequations}
\label{eq17h}
\begin{eqnarray}
\label{eq17-h}
N<&&B(n,p,p')\left[\int_0^{\infty} dr\, r^{2(p-1)/p}\, |V^-(r)|\right]\left[\int_0^{\infty} dr\, 
|V^-(r)|^{pp'}\right]^{1/(pp')}
\nonumber \\
&&\left[\int_0^{\infty} dr\, r^{(p'-1)/p'}\, |V^-(r)|^p \right]^{1/p}
\, \left[\int_0^{\infty} dr\, r^{p-1}\, |V^-(r)|^p \right]^{(n-3)/p},
\end{eqnarray}
with the constant $B(n,p,p')$ given by
\begin{equation}
\label{eq17-i}
B(n,p,p')=\sum_{\nu=0}^{\infty} (2\nu+1)\, [C(\nu,p/(p-1))]^{n-1}\, C(\nu,pp'/(p'-1)),
\end{equation}
and
\begin{eqnarray}
\label{eq17-j}
C(\nu,q)=&&\frac{1}{\alpha \sqrt{\pi}}\frac{\Gamma(\nu+1)}{2^{2\nu+2}\Gamma(\nu+3/2)}\nonumber \\
&& \left[\int_0^1 dx\, (1+1/x^2)\left(z^{\nu+1}F(\nu+1,\nu+1,2(\nu+1),z)\right)^q\right]^{1/q},
\end{eqnarray}
\end{subequations}
with $z=4x/(1+x)^2$, $p>1$ and $p'>1$. All the complexity of the problem is now located in the calculation of the constant $B(n,p,p')$. Analytical calculations for small values of $\nu$ and numerical investigations up to $\nu=100$ seems to prove that for all values of $\nu$ we have
\begin{subequations}
\label{eq17k}
\begin{equation}
\label{eq17-k}
C(\nu,2)=1/(\alpha\sqrt{2\nu+1}),
\end{equation}
\begin{equation}
\label{eq17-kb}
C(\nu,3)\le 1/(\alpha(2\nu+1)^{1/3}).
\end{equation}
\end{subequations}
In this case, we find that 
$B(n,2,3)\le (1-2^{-w})\zeta(w)$, with $w=(3n-7)/6$ and $\zeta(x)$ is the Riemann Zeta function. This implies $n\ge 5$ to obtain nontrivial results and in particular $B(5,2,3)\le 2.172$.

Daubechies has obtained the following simple upper limit on the total number of bound states \cite{daub83}
\begin{equation}
\label{eq17-l}
N\le K \int d{\bf r}\, [|V^-(r)|(|V^-(r)|+2m)]^{3/2},
\end{equation}
with $K=0.239$ for arbitrary values of $m$ and $K=0.103$ for $m=0$. This inequality shows that $N$ grows with strength of the potential, $g$, at most as $g^3$. The upper limits obtained in this section behave as $g^n$, with $n\ge 4$, and should not yield very cogent results for large value of $g$ when the potential possesses many bound states. Nevertheless, in the Section \ref{subsec3.3}, we obtain an upper limit on the total number of bound states for central potentials, and $m=0$, which features the correct dependency on $g$. 

\subsection{Upper limit on the number of $\ell$-wave bound states}
\label{subsec3.2}

To obtain an upper limit on the number of $\ell$-wave bound states of the spinless Salpeter equation, we need to derive the radial version of the integral equation (\ref{eq17}). This can be achieved by performing similar calculations than those done in Ref. \cite[p. 2255-2257]{brau98}. Integration over angular variables reduces the integral equation (\ref{eq16}) to the following one-dimensional integral equation
\begin{equation}
\label{eq18}
u_{\ell}(r)=\int_0^{\infty} dr'\, G_{\ell}(m,r,r')\, |V^-(r')|\, u_{\ell}(r'),
\end{equation}
with
\begin{equation}
\label{eq19}
G_{\ell}(m,r,r')=\frac{mrr'}{2}\int_0^{\pi}d\theta' \, \sin \theta' \, \frac{H(m\Delta)}{\Delta}\, P_{\ell}(\cos \theta'),
\end{equation}
where $u_{\ell}(r)$ is the radial wave function, 
$\Psi({\bf r})=(u_{\ell}(r)/r)Y_{\ell m}({\bf \hat{r}})$ and $H(x)$ is defined by 
(\ref{eq15a}). As we pointed out in Section \ref{sec2}, the integral equation (\ref{eq18}) can be written with a symmetrical kernel provided we introduce a new wave function
\begin{equation}
\label{eq20}
\phi_{\ell}(r)=|V^-(r)|^{1/2}\, u_{\ell}(r).
\end{equation}
This change of function leads to the following integral equation
\begin{equation}
\label{eq21}
\phi_{\ell}(r)=\int_0^{\infty}dr'\, K_{\ell}(m,r,r')\, \phi_{\ell}(r'),
\end{equation}
with
\begin{equation}
\label{eq22}
K_{\ell}(m,r,r')=|V^-(r)|^{1/2}\, G_{\ell}(m,r,r')\,|V^-(r')|^{1/2}.
\end{equation}

We can now again use the Birman-Schwinger method, see Section {\ref{sec2}, with the integral equation (\ref{eq21}) to obtain an upper limit on the number of $\ell$-wave bound states. The relation (\ref{eq11}) together with the majorization (\ref{eq17-c}) lead to 
\begin{subequations}
\label{eq24}
\begin{equation}
\label{eq24a}
N_{\ell}<\int_0^{\infty} dr_1 \ldots dr_n\, |V^-(r_1)|\ldots |V^-(r_n)|\, T_{\ell}(m,r_1,r_2)\ldots T_{\ell}(m,r_n,r_1),
\end{equation}
with
\begin{equation}
\label{eq24b}
\alpha\, T_{\ell}(m,r,r')=\frac{1}{\pi}{\cal G}_{\ell}(m,r,r')+{\cal S}_{\ell}(m,r,r'),
\end{equation}
and
\begin{equation}
\label{eq24c}
{\cal G}_{\ell}(m,r,r')=m\int_{|r-r'|}^{r+r'}dy\, K_1(my)\, P_{\ell}\left(\frac{r^2+r'^2-y^2}{2rr'}\right),
\end{equation}
\begin{equation}
\label{eq24d}
{\cal S}_{\ell}(m,r,r')=m\int_{|r-r'|}^{r+r'}dy\, P_{\ell}\left(\frac{r^2+r'^2-y^2}{2rr'}\right)=
\frac{2m}{2\ell+1}\, r_<^{\ell+1}\, r_>^{-\ell},
\end{equation}
\end{subequations}
where $r_<=\min[r,r']$ and $r_>=\max[r,r']$. The kernel ${\cal S}_{\ell}(m,r,r')$ is actually the Green function of the nonrelativistic kinetic energy operator and takes a simple form while the kernel ${\cal G}_{\ell}(m,r,r')$ can be calculated analytically for each value of $\ell$ \cite{nick84,brau98}. We have, here also, introduced in (\ref{eq24b}) the parameter $\alpha$ which takes the value 1 respectively 2 for one respectively two (identical) particle problems. We can use an additional approximation and majorize $K_1(y)$ by $1/y$. In this case, $T_{\ell}(m,r,r')$ is replaced by $A_{\ell}(m,r,r')$, see (\ref{eq17-f}). Note that $T_{\ell}(0,r,r')=A_{\ell}(0,r,r')$ since in this limit ($m\rightarrow0$) $m K_1(my)=1/y$.

As indicated in the previous Section \ref{subsec3.1}, simplifications occur in the ultrarelativistic case ($m=0$). With the help of the H\"older inequality we obtain
\begin{subequations}
\label{eq25}
\begin{eqnarray}
\label{eq25a}
N_{\ell}<&&\tilde{B}(n,\ell,p,p')\left[\int_0^{\infty} dr\, r^{2(p-1)/p}\, |V^-(r)|\right]
\left[\int_0^{\infty} dr\, |V^-(r)|^{pp'}\right]^{1/(pp')}\nonumber \\
&&\left[\int_0^{\infty} dr\, r^{(p'-1)/p'}\, |V^-(r)|^p \right]^{1/p}
\left[\int_0^{\infty} dr\, r^{p-1}\, |V^-(r)|^p \right]^{(n-3)/p},
\end{eqnarray}
with 
\begin{equation}
\label{eq25b}
\tilde{B}(n,\ell,p,p')=[C(\ell,p/(p-1))]^{n-1}\, C(\ell,pp'/(p'-1)),
\end{equation}
\end{subequations}
and where $n\ge 2$, $p>1$, $p'>1$ and $C(\ell,q)$ defined by (\ref{eq17-j}). The upper limit on the total number of bound states that can be obtained from the upper limit on the number of $\ell$-wave bound states (\ref{eq25}) is very similar to the upper limit (\ref{eq17-h}) except that the sum would stop at $\ell=L$ (the largest value of the angular momentum $\ell$ for which bound states do exist) and that $n$ is only restricted to be greater than two. 

Taking the limit $n\rightarrow \infty$, we obtain the following necessary condition for the existence of at least one $\ell$-wave bound state:
\begin{equation}
\label{eq25-b}
\int_0^{\infty} \frac{dr}{r}\, \left[C(\ell,p/(p-1))\, r\, |V^-(r)|\right]^p \ge 1.
\end{equation}
This simple relation yields a lower limit, $g_{\text{c}}$, on the (numerically) exact ``critical" value of $g$, $g_{\text{c,ex}}$, for which a first bound state appears, 
$g_{\text{c,ex}}\ge g_{\text{c}}$ \textit{and} yields an upper limit $L^+$ on $L$. We can obtain a simpler expression by considering now the limit $p\rightarrow \infty$. We find that a necessary condition for the existence of at least one $\ell$-wave bound state is also given by
\begin{subequations}
\label{eq25c}
\begin{equation}
\label{eq25-c}
C(\ell,1)\, {\cal M} \ge 1,
\end{equation}
where
\begin{equation}
\label{eq25-d}
{\cal M} = \max[r\, |V^-(r)|].
\end{equation}
\end{subequations}
This last necessary condition have only a sens for $\ell >0$ because $C(0,1)$ diverges. Of course, this does not mean that there always exist bound states for $\ell=0$. For this value of the angular momentum, one needs to use the relation (\ref{eq25-b}) to draw conclusions about existence of bound states. The constant $C(\ell,1)$ can be rewritten as
\begin{equation}
\label{eq25-e}
C(\ell,1)=\frac{1}{\alpha \sqrt{\pi}}\frac{\Gamma(\ell+1)}{\Gamma(\ell+3/2)} \frac{c(\ell)}{\sqrt{2\ell+1}}.
\end{equation}
The interest of this rewritten is that the function $c(\ell)$ varies very slowly with $\ell$ and can then be easily and usefully tabulated, see Table~\ref{tab1}.
A possible simple majorization of $c(\ell)$ is $c(\ell)< \sqrt{2\pi}+a \ell^b$ with $a= 0.7$ and $b= -1.18$ 
(valid at least for $1\le \ell \le 100$). Note also that the necessary condition (\ref{eq25c}) is the ultrarelativistic counterpart of the well known nonrelativistic necessary condition 
$(\ell+1/2)^{-2} \max[r^2 \, |2mV^-(r)|]\ge 1$ (see for example \cite{cour53}).

\begin{table}
\protect\caption{Values of the function $c(\ell)$, see (\protect\ref{eq25-e}), for several values of $\ell$. Note that $c(0)=\infty$ and $c(\infty)=\sqrt{2\pi}$.}
\label{tab1}
\begin{center}
\begin{tabular}{|c|c|c|c|c|c|c|c|}
\hline
$\ell$ & $c(\ell)$ & $\ell$ & $c(\ell)$ & $\ell$ & $c(\ell)$ & $\ell$ & $c(\ell)$  \\
\hline
1 &  3.205  &  8  &  2.557   &  15  &  2.531  &  40  &  2.515            \\
2 &  2.795  &  9  &  2.550   &  16  &  2.529  &  50  &  2.513          \\
3 &  2.678  & 10  &  2.545   &  17  &  2.528  &  60  &  2.512       \\
4 &  2.625  & 11  &  2.541   &  18  &  2.526  &  70  &  2.511         \\
5 &  2.596  & 12  &  2.538   &  19  &  2.525  &  80  &  2.511        \\
6 &  2.578  & 13  &  2.535   &  20  &  2.524  &  90  &  2.510         \\
7 &  2.566  & 14  &  2.533   &  30  &  2.518  & 100  &  2.510         \\
\hline
\end{tabular}
\end{center}
\end{table}
 
\subsection{Upper limit on the total number of bound states for central potentials in the ultrarelativistic regime}
\label{subsec3.3}

In this section, we use the upper limit (\ref{eq25}) obtained in the previous section to infer a limit on the total number of bound states applicable only to central potentials in the ultrarelativistic regime. We choose in (\ref{eq25}) $n=2$, $p=2$ and $p'\rightarrow \infty$ and with the help of (\ref{eq17-k}) (which is certainly correct up to $\ell=100$) we obtain
\begin{subequations}
\label{eq25f}
\begin{equation}
\label{eq25-f}
N_{\ell}<\frac{1}{\alpha^2 (2\ell+1)}\,\tilde{{\cal M}}\,{\cal I} ,
\end{equation}
with
\begin{equation}
\label{eq25-g}
\tilde{{\cal M}}=\max[|V^-(r)|],
\end{equation}
\begin{equation}
\label{eq25-h}
{\cal I}=\int_0^{\infty} dr\, r\, |V^-(r)|.
\end{equation}
\end{subequations}
To some extend, the limit (\ref{eq25f}) is the ultrarelativistic counterpart of the nonrelativistic Bargmann-Schwinger upper bound \cite{ba52,sc61}. The upper limit on $N$ is obtained with a summation over the right-hand side of (\ref{eq25-f}) from $\ell=0$ to $\ell=L$ taking into account the multiplicity of each $\ell$-wave bound states. To this end, an upper limit on $L$ is needed. The best limit, which behaves linearly with the strength of potential, is obtained, not from (\ref{eq25-f}), but instead from the simple relation (\ref{eq25c}). We have 
\begin{equation}
\label{eq25-i}
N<\frac{1}{\alpha^2}(L^+ +1)\,\tilde{{\cal M}}\,{\cal I}.
\end{equation}
The optimal value for $L^+$ is obtained by solving (\ref{eq25c}) but a neater, if generally less stringent, upper limit $L^{++}$ ($L^+\le L^{++}$) is given by
\begin{subequations}
\label{eq25j}
\begin{equation}
\label{eq25-j}
L^{++} = \frac{1}{4}\left(\sqrt{1+8s^2}-1\right),
\end{equation}
with
\begin{equation}
\label{eq25-k}
s=\frac{c(1)}{\alpha\sqrt{\pi}}\, {\cal M}.
\end{equation}
\end{subequations}
The expression (\ref{eq25j}) for the upper limit on $L$ is obtained using (\ref{eq25c}), (\ref{eq25-e}) and the second part of the inequalities $\sqrt{x+1}\ge \Gamma[x+3/2]/\Gamma[x+1]\ge \sqrt{x}$ for $x\ge 0$. $c(1)$ could even be replaced by $c(L^{++})$ in (\ref{eq25-k}). The transcendental equation obtained is solved very quickly after few iterations thanks to the slow variation of $c(\ell)$ as a function of $\ell$. If we believe that the relation (\ref{eq17-k}) is always true for all values of the angular momentum, we can write the asymptotic expression of the upper limit (\ref{eq25-i}) when the strength of the potential, $g$, goes to infinity:
\begin{equation}
\label{eq25-l}
N(g\rightarrow \infty)<\frac{1}{\alpha^3}{\cal M}\,\tilde{{\cal M}}\,{\cal I},
\end{equation}
where $c(\infty)$ were used to obtain the asymptotic expression of $L^+$, $L^+\approx{\cal M}/\alpha$, where the symbol $\approx$ means asymptotic equality. This last expression is useful to compare the upper limit (\ref{eq25-i}) with the Daubechies upper limit (\ref{eq17-l}) since we clearly have
\begin{equation}
\label{eq25-m}
\int_0^{\infty} dr \, r^2\, |V^-(r)|^3\le {\cal M}\,\tilde{{\cal M}}\,{\cal I}.
\end{equation}
This last inequality means that the upper limit (\ref{eq17-l}) would always be better than the limit (\ref{eq25-i}) if its coefficient was equal to 1. Since the coefficient is greater than unity, $4\pi K= 1.294>1$, there is some room for the limit (\ref{eq25-i}) to be more stringent.
A square well potential is an example:
\begin{equation}
\label{eq25-n}
V(r)=-V_0\, \theta[(R_1-r)(r-R_2)],
\end{equation}
where $R_2\ge R_1$ are two arbitrary positive radius and $\theta(x)$ is the step function ($\theta(x)=1$ if $x\ge 0$ and $\theta(x)=0$ if $x<0$). When the ratio of the radius, $R_1/R_2$,  is in the interval $[0.4859,1)$, the upper limit (\ref{eq25-i}) is better than the Daubechies upper limit.
  
\section{Tests}
\label{sec4}

In this section, we test the limits obtained in the previous sections with two simple central potentials written hereafter as $V(r)=-gR^{-1}v(r/R)$. This first one is an exponential potential
\begin{equation}
\label{eq28}
v(x)=\exp(-x).
\end{equation}
The second one is a P\"oschl-Teller potential
\begin{equation}
\label{eq29}
v(x)=\frac{1}{\cosh^2{x}}.
\end{equation}

The upper limits (\ref{eq17e}) and (\ref{eq17h}) on the total number of bound states behave with the strength of the potential, $g$, at least as $g^4$ and should not be very stringent for strong potential possessing many bound states. Nevertheless, for weak potentials these upper bounds could be rather effective, especially to establish necessary condition for the existence of at least one bound state. The same remark applies to the upper limits (\ref{eq24}) and (\ref{eq25}) since they behave at least as $g^2$ instead of $g$.

\subsection{The S-wave case}
\label{subsec4.1}

For simplicity, we restrict our attention to the $n=2$ case to test the upper limit (\ref{eq24}) in the S-wave case. This limit, for $\ell=0$, reads
\begin{eqnarray}
\label{eq26}
N_0&<&\frac{1}{(\pi\alpha)^2}g^2\int_0^{\infty} dx\,v(x)\int_0^{\infty}dy\, v(y)\, \big[
(K_0(\beta|x-y|)-K_0(\beta(x+y)))\nonumber \\ &+&\pi\beta(x+y-|x-y|)\big]^2,
\end{eqnarray}
with $\beta=mR$. $g$ and $\beta$ are two dimensionless parameters. While the dependency of (\ref{eq26}) on $g$ is very simple, the behavior of this limit with a variation of $\beta$ is more involved (except for large $\beta$, see below). Simplifications obviously occur in the case of a vanishing mass and the limit reads
\begin{equation}
\label{eq27}
N_0<\frac{1}{(\pi\alpha)^2}g^2\int_0^{\infty} dx\,v(x)\int_0^{\infty}dy\, v(y)\, 
\log^2\left|\frac{x+y}{x-y}\right|.
\end{equation}

We test the above limits by computing the necessary condition on $g$, $g\ge g_{\text{c}}$, for the existence of at least one bound state. In other words, we calculate a lower limit, $g_{\text{c}}$, on the (numerically) exact ``critical" value of $g$, $g_{\text{c,ex}}$, for which a first bound state appears, $g_{\text{c,ex}}\ge g_{\text{c}}$. The numerical calculations are performed with the help of the very accurate Lagrange Mesh method \cite{baye86}. The comparison between the (numerically) exact results, $g_{\text{c,ex}}$, and the lower limits, $g_{\text{c}}$, is displayed in Table~\ref{tab2} for a 
two-particles problem ($\alpha=2$). We choose a two identical particles problem only for numerical convenience (to not modify numerical codes). The results obtained with the upper limits (\ref{eq26}) and (\ref{eq27}) are rather satisfactory especially when $\beta$ is large and are always better than the results obtained with the Daubechies upper limit 
(\ref{eq17-l}). 

\begin{table}
\protect\caption{Comparison, for the exponential and the P\"oschl-Teller potentials, between the critical values, $g_{\text{c}}$, of $g$ yielded by the limits (\protect\ref{eq26}), (\protect\ref{eq27}), the critical values, $g_{\text{c,D}}$, of $g$ yielded by the Daubechies upper limit (\protect\ref{eq17-l}) and the exact critical value $g_{\text{c,ex}}$, obtained by solving numerically the spinless Salpeter equation.}
\label{tab2}
\begin{center}
\begin{tabular}{|c|c|c|c|c|c|c|}
\hline
& \multicolumn{3}{c|}{Exponential} & \multicolumn{3}{c|}{P\"oschl-Teller}  \\
\cline{2-7}
$\beta$  &$g_{\text{c}}$ &$g_{\text{c,D}}$ & $g_{\text{c,ex}}$ & $g_{\text{c}}$ &$g_{\text{c,D}}$& $g_{\text{c,ex}}$\\
\hline
0 &  4.443  & 4.370  &  5.574   &   4.126  & 3.886  &  5.008    \\
1 &  1.223  & 0.6574 &  1.361   &   1.512  & 0.8631 &  1.742    \\
2 &  0.6739 & 0.3374 &  0.7133  &   0.8912 & 0.4582 &  0.9598   \\
3 &  0.4604 & 0.2261 &  0.4804  &   0.6233 & 0.3092 &  0.6549   \\
4 &  0.3487 & 0.1698 &  0.3616  &   0.4769 & 0.2329 &  0.4956   \\
5 &  0.2803 & 0.1360 &  0.2898  &   0.3854 & 0.1867 &  0.3981   \\
\hline
\end{tabular}
\end{center}
\end{table}

In the limit of large $\beta$, only the kernel ${\cal S}_0(m,r,r')$ will obviously contribute significantly. In this asymptotic regime, the lower limit $g_{\text{c}}$ takes a very simple form, and reads for the exponential (E) and the P\"oschl-Teller (PT) potentials 
\begin{equation}
\label{eq30}
g_{\text{c}}^{\text{E}}\approx\frac{\sqrt{2}}{\beta}=\frac{1.4142}{\beta} \quad \text{and} \quad g_{\text{c}}^{\text{PT}}\approx\frac{1.9663}{\beta}, 
\end{equation}
where the symbol $\approx$ means asymptotic equality.
But the limit of large $\beta$ is equivalent to the limit of large $m$, which leads to nonrelativistic regime. We can thus compare the lower limits (\ref{eq30}) with those obtained from nonrelativistic upper bounds on the number of bound states, and also with the nonrelativistic exact results. The most stringent nonrelativistic lower bound $g_{\text{c,NR}}$ is obtained with the upper limit found in \cite{gl76} (see also \cite{ch95b}) and reads
\begin{equation}
\label{eq31}
g_{\text{c,NR}}^{\text{E}}=\frac{1.4383}{\beta}\quad \text{and} \quad g_{\text{c,NR}}^{\text{PT}}=\frac{1.9910}{\beta}. 
\end{equation}
The exact result, for the exponential potential, is obtained in terms of the first zero, $z_0=2.4048$, of the Bessel function $J_0(x)$ (see for example \cite[p. 196]{flug94}) while the exact result, for the P\"oschl-Teller potential, is extracted from the exactly known spectra (see for example \cite[p. 94]{flug94}). These exact results read
\begin{equation}
\label{eq32}
g_{\text{c,NR,ex}}^{\text{E}}=\frac{z_0^2}{4\beta}=\frac{1.4458}{\beta}\quad \text{and} \quad g_{\text{c,NR,ex}}^{\text{PT}}=\frac{2}{\beta}. 
\end{equation}
At first sight, it is surprising that the upper limit on the number of bound states derived for spinless Salpeter equation (which reduces to the Schr\"odinger equation in the limit of large $m$) leads asymptotically to similar restrictions on $g$, for the existence of at least one bound state, than previously known nonrelativistic upper limits. It should be noted that the limits obtained in this article has no counter part in the nonrelativistic context since we have used iterated kernels to obtain these upper limits, and in particular the limit (\ref{eq24}). Indeed, it is not necessary to use iterated kernels to obtain an upper limit on the number of $\ell$-wave bound states in the context of nonrelativistic quantum mechanics (the Bargmann-Schwinger upper limit \cite{ba52,sc61}) since the kernel is not singular. Moreover, the use of an unnecessary iterated kernel would lead to an even more incorrect dependency on $g$ (the Bargmann-Schwinger upper limit does not display the correct dependency on $g$, see for example \cite{ca65a}), which would lead to a poorer upper limit for a strong potential possessing many bound states. Nevertheless, it seems that such a poor upper limit could yield very strong conditions on $g$ for the absence of bound states. This remark and the potential benefit which could be obtained from it, namely obtaining strong conditions for the absence of bound states in the context of nonrelativistic quantum mechanics, will be studied in detail elsewhere.

These comparisons with nonrelativistic results and those reported in the Table~\ref{tab2} show that the upper limit (\ref{eq24}) is rather effective to establish necessary condition for the existence of at least one bound state for the spinless Salpeter equation, especially when $\beta$ becomes large. This also indicates that this upper limit should yield cogent information for weak potentials possessing few bound states. The results presented in Table~\ref{tab2}, for $n=2$, should be improved as $n$ goes to infinity and, we believe, should converge to the exact result if there were not the majorization (\ref{eq17-c}).

We can also test the upper limit (\ref{eq25}) and the necessary condition (\ref{eq25-b}) (both valid only when $m=0$) by the computation of $g_{\text{c}}$. The series of $g_{\text{c}}$ obtained with (\ref{eq25}) for each value of $n$ are monotonic series which converges to $g_{\text{c}}=4$ for the exponential potential and converges to $g_{\text{c}}= 3.685$ for the P\"oschl-Teller potential, which are, obviously, the values obtained with 
the expression (\ref{eq25-b}).

\subsection{The $\ell$-wave case}
\label{subsec4.2}

In this section we only test the relations (\ref{eq25-b}) and (\ref{eq25c}) by computing the lower limit $g_{\text{c}}$ for $\ell>0$ with the potentials (\ref{eq28}) and (\ref{eq29}). The comparison of these lower limit with the (numerically) exact critical coupling constant is given in the Table~\ref{tab3}. 

\begin{table}
\protect\caption{Comparison, for the exponential and the P\"oschl-Teller potentials, between the critical values, $g_{\text{c}}$, of $g$ yielded by the limits (\protect\ref{eq25-b}) ($g_{\text{c,I}}$) and (\protect\ref{eq25c}) ($g_{\text{c,II}}$) and the exact critical value $g_{\text{c,ex}}$, obtained by solving numerically the spinless Salpeter equation.}
\label{tab3}
\begin{center}
\begin{tabular}{|c|c|c|c|c|c|c|}
\hline
& \multicolumn{3}{c|}{Exponential} & \multicolumn{3}{c|}{P\"oschl-Teller}  \\
\cline{2-7}
$\ell$  &$g_{\text{c,I}}$ &$g_{\text{c,II}}$ & $g_{\text{c,ex}}$ & $g_{\text{c,I}}$ &$g_{\text{c,II}}$& $g_{\text{c,ex}}$\\
\hline
1 &  8.524 & 6.922 & 10.98 &   7.437 & 5.687 & 9.545  \\
2 &  13.67 & 12.81 & 16.39 &   11.59 & 10.53 & 14.04  \\
3 &  19.03 & 18.46 & 21.81 &   15.91 & 15.17 & 18.52  \\
4 &  24.44 & 24.02 & 27.24 &   20.30 & 19.73 & 22.99  \\
5 &  29.88 & 29.53 & 32.67 &   24.73 & 24.27 & 27.46  \\
\hline
\end{tabular}
\end{center}
\end{table}

The lower limits yielded by the relation (\ref{eq25-b}) are always better than those obtained with the relation (\ref{eq25c}) but the differences become smaller as $\ell$ grows. These lower limits are quite satisfactory compared to the exact results and the relative differences between these quantities decrease from $22\%$ to $10\%$ for $\ell$ increasing from 1 to 5 for both potentials.

\subsection{Tests of an upper limit on $L$}
\label{subsec4.3}

To conclude this section devoted to test the limits obtained in this article, we compare the upper limit, $L^+$, obtained with (\ref{eq25c}), with the exact largest value, $L$, of the angular momentum for which bound states do exist. The results obtained for an exponential and a P\"oschl-Teller potential are reported in the Table~\ref{tab4}. The bounds on $L$ obtained with (\ref{eq25c}) are very stringent for both potentials. These excellent results are not so surprising since the same strong limitations on $L$ are obtained with the nonrelativistic counterpart of (\ref{eq25c}).

\begin{table}
\protect\caption{Comparison, for the exponential and the P\"oschl-Teller potentials, between the upper limit, $L^+$, obtained with (\ref{eq25c}), with the exact largest value, $L$, (obtained by solving numerically the spinless Salpeter equation) of the angular momentum for which bound states do exist.}
\label{tab4}
\begin{center}
\begin{tabular}{|c|c|c|c|c|}
\hline
& \multicolumn{2}{c|}{Exponential} & \multicolumn{2}{c|}{P\"oschl-Teller}  \\
\cline{2-5}
 $g$ &$L^+$ &$L$ & $L^+$ & $L$ \\
\hline
10  &  1  & 0 &  1  & 1  \\
20  &  3  & 2 &  4  & 3  \\
30  &  5  & 4 &  6  & 5  \\
40  &  6  & 6 &  8  & 7  \\
50  &  8  & 8 &  10 & 10 \\
100 &  17 & 17&  21 & 21 \\
150 &  27 & 26&  33 & 32 \\
200 &  36 & 35&  44 & 43 \\
\hline
\end{tabular}
\end{center}
\end{table}

\section{Upper limit on the number of bound states lying below a given energy}
\label{sec5}
 
To conclude this letter, it is worth mentioning that the method used to obtain the upper limits presented in this work can also be used to derive upper limits on the number of states of a given potential that lie below a given energy. Indeed, instead of using 
$-|V^-(r)|$ as a comparison potential, which admit at least as much bound states as $V(r)$, we can use a potential $\tilde{V}(r)$ that equals $V(r)$ wherever $V(r)<-\kappa^2$ ($\kappa^2$ is an arbitrary real energy) and equals $-\kappa^2$ in the regions for which $V(r)\ge-\kappa^2$. But this latter potential, $\tilde{V}(r)$, is equivalent to the potential $V_{\kappa}(r)$ that equals $V(r)+\kappa^2$ wherever this quantity is negative and equals zero otherwise, since $V_{\kappa}(r)=\tilde{V}(r)+\kappa^2$. Thus, an upper bound to the total number of $\ell$-wave bound states associated with $V_{\kappa}(r)$ serves to limit the number of $\ell$-wave states that lie at or below the energy $E=-\kappa^2$ for the potential $V(r)$. For example, the limit (\ref{eq24}) for $n=2$ reads
\begin{equation}
\label{eq33}
N_{\ell}(E\le-\kappa^2)<\int_0^{\infty} dr\, V_{\kappa}(r)\int_0^{\infty}dr'\, V_{\kappa}(r')\, T^2_{\ell}(m,r,r')\Big]_{V+\kappa^2<0},
\end{equation}
with $T_{\ell}(m,r,r')$ defined by (\ref{eq24b}). This last upper limit can be tested with the harmonic oscillator potential, $V(r)=k^3 r^2$, with $m=0$ (with a slight adaptation since $V(r)$ is now positive). Indeed, in this case the spectrum is exactly known for $\ell=0$ \cite{luch91}:
\begin{equation}
\label{eq34}
E_n=k\, \lambda_n,
\end{equation}
where $-\lambda_n$ are the zeros of the Airy function. Thus, below these eigenenergies, there are exactly $n-1$ bound states. For $n=1$, 2 and 3, the upper limit gives 1, 8, 21. As expected, the limit gives cogent results when few bound states exist and becomes less stringent as the number of bound states grows because of the incorrect behavior with the growth of the coupling constant. 

{\bf Acknowledgments}

We would like to thank Prof. F. Calogero for reading the manuscript.

\end{document}